\documentclass[epj]{webofc}
\usepackage[varg]{txfonts}   
\woctitle{13th International Conference on Meson-Nucleon Physics and the Structure of the Nucleon (MENU 2013)}
\begin{document}
\title{$\pi N$ TDAs from charmonium production in association with a forward pion at \={P}ANDA}

\author{B.~Ma \inst{1}
\and
    B.~Pire\inst{2}
    \and
        K.~Semenov-Tian-Shansky\inst{3}\fnsep\thanks{\email{ksemenov@ulg.ac.be}} \and
        L.~Szymanowski\inst{4}
}

\institute{
IPN, Universit\'{e} Paris Sud, IN2P3/CNRS, F-91406 Orsay, France 
\and
CPhT, {\'E}cole Polytechnique, CNRS, F-91128 Palaiseau, France
\and
IFPA, d\'{e}partement AGO,  Universit\'{e} de  Li\`{e}ge, 4000 Li\`{e}ge,  Belgium
\and
National Centre for Nuclear Research (NCBJ), Warsaw, Poland
          }

\abstract{%
  Charmonium production in association with a forward or backward pion in nucleon-antinucleon annihilation
  is one of the most promising reaction to access nucleon-to-pion transition distribution amplitudes (TDAs)
  at \={P}ANDA. We briefly review the description of this reaction in terms of 
  $\pi N$ 
  TDAs within the collinear factorization approach   and present the first results of dedicated feasibility studies for 
  the \={P}ANDA  experimental setup.
}
\maketitle
\section{Introduction}
\label{intro_semenov}
Baryon-to-meson transition distribution amplitudes (TDAs) 
\cite{TDAfirst,Lansberg:2007ec,Lansberg:2007se,Pire:2010if,Pire:2011xv,Lansberg:2011aa,Lansberg:2012ha} 
are universal non-perturbative objects which share common features both with generalized parton distributions 
(GPDs) and baryon distribution amplitudes (DAs) and arise in the collinear factorized description of several 
classes of hard exclusive reactions.

Dedicated studies of hard exclusive reactions providing access to
baryon-to-meson TDAs were suggested
\cite{Wiedner:2011mf}
as a part of the future experimental program of \={P}ANDA facility
\cite{PANDAprogramm},
which is currently under construction at  GSI-FAIR.
Extracting baryon-to-meson TDAs from the experimental measurements will
provide  valuable complementary information on the hadronic structure  and
deepen our understanding of the strong interaction phenomena (see {\it e.g.} the discussion in
refs.~\cite{Pire:2011xv,Lansberg:2011aa}) .

The \={P}ANDA experimental setup provides several options to access hard exclusive reactions
admitting description in terms of baryon-to-meson TDAs. Prominent examples are nucleon-antinucleon
annihilation into the high invariant mass lepton pair
\cite{Lansberg:2007se,Lansberg:2012ha}
(or charmonium \cite{Pire:2013jva}) 
in association with a pion.
Detailed feasibility studies of the corresponding reactions are of major importance since
measurements at \={P}ANDA
will give the unique
opportunity to check explicitly the universality
\cite{Muller:2012yq}
of baryon-to-meson TDAs
providing information on the time-like regime, which is complementary to that from the lepton beam induced
reactions, which can be studied {\it e.g.} at JLab
\cite{Lansberg:2007ec,Lansberg:2011aa}.

\section{QCD description of charmonium plus pion production in
antinucleon-nucleon annihilation}
\label{sec-1_semenov}

\begin{figure}
\centering
\includegraphics[width=4.5cm,clip]{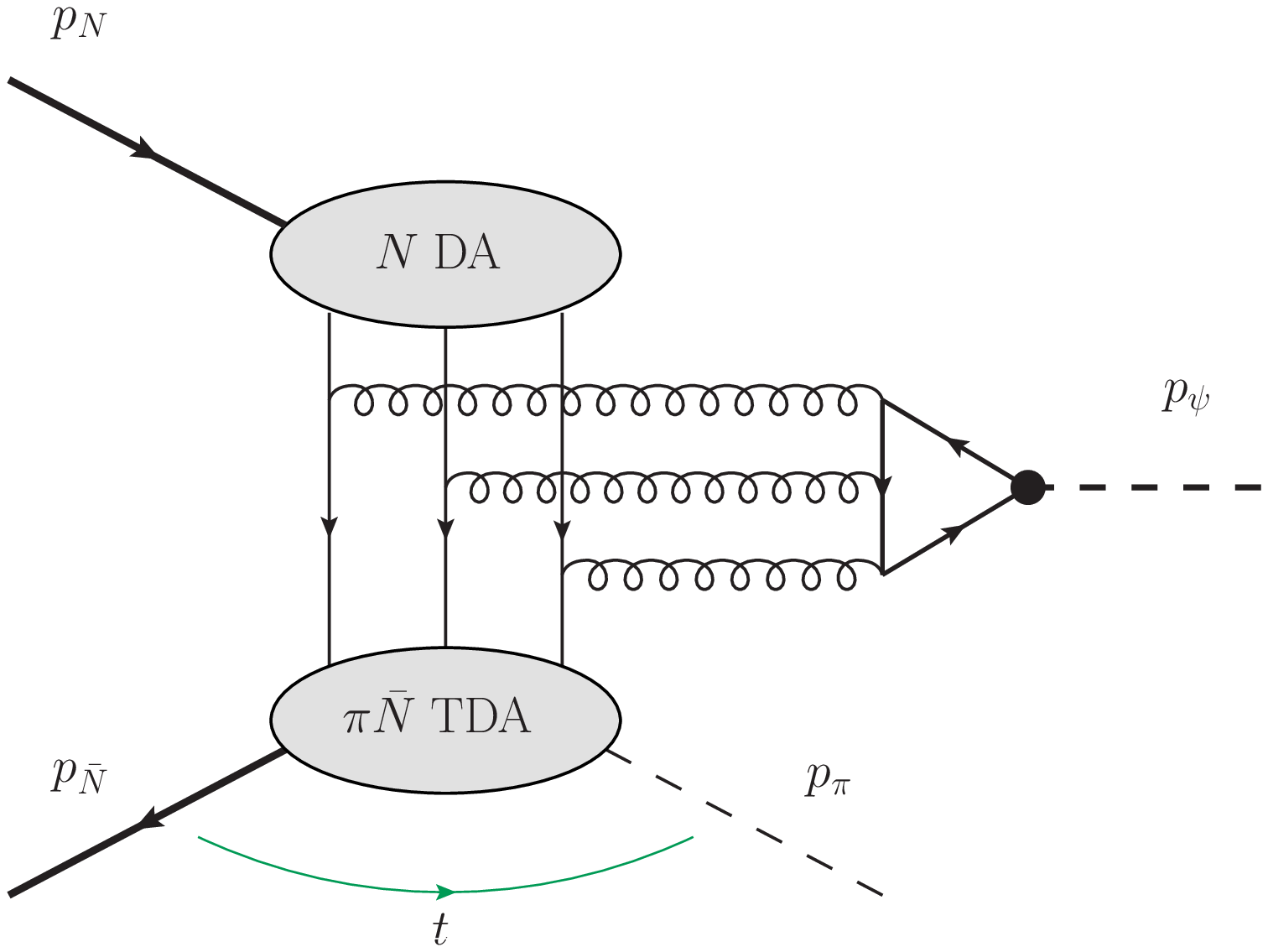}
\includegraphics[width=4.5cm,clip]{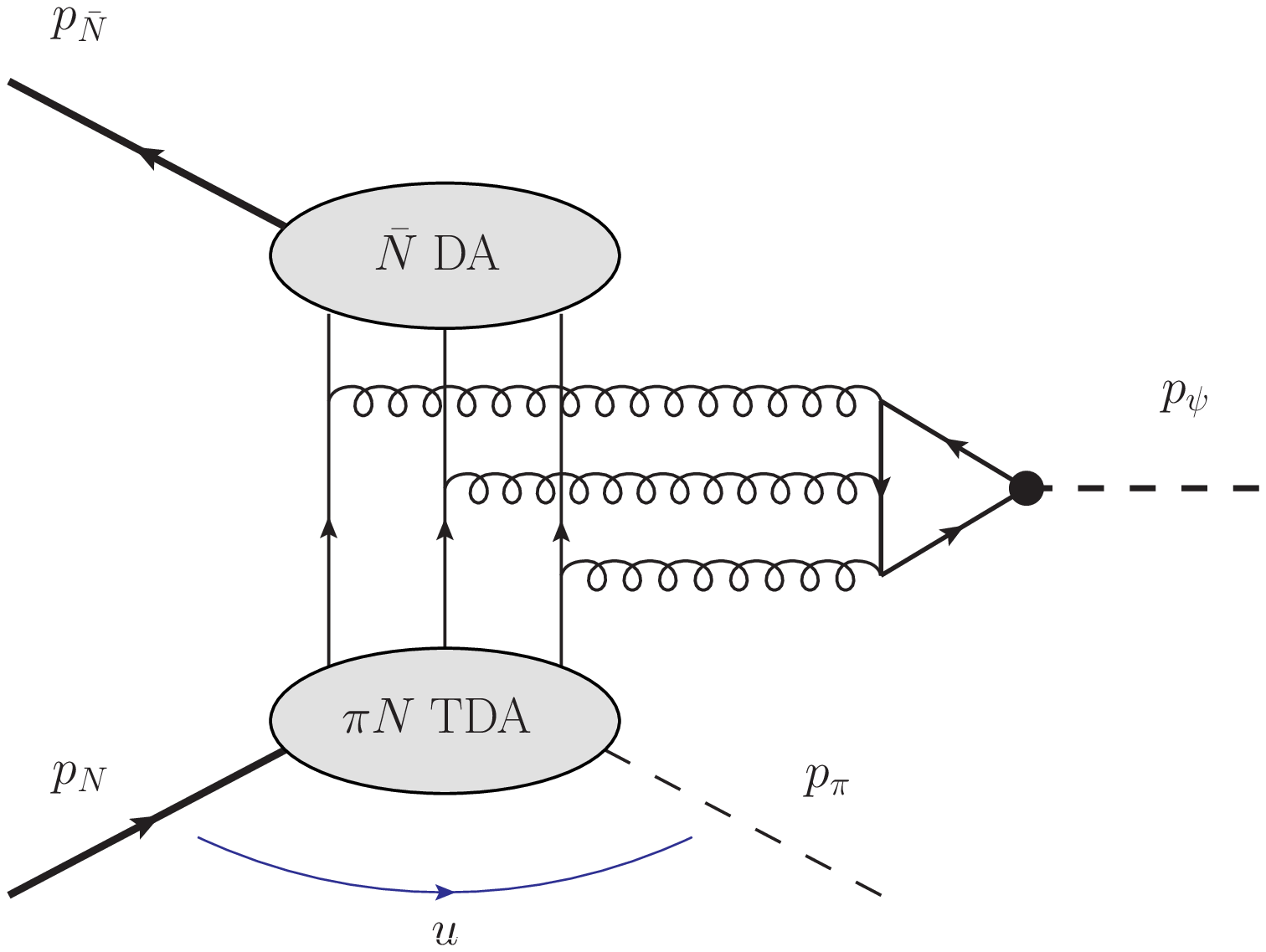}
\caption{Collinear factorization of the annihilation process
(\ref{reac}).
{\bf Left panel:} forward kinematics
($t \sim 0$).
{\bf Right panel:} backward kinematics
($u \sim 0$).
$N(\bar{N})$
DA stands for the distribution amplitude of antinucleon (nucleon);
$\pi \bar{N} (\pi N)$
TDA stands for the transition distribution amplitude from an antinucleon (nucleon) to a pion.}
\label{fig-0_Semenov}       
\end{figure}

Study of charmonium production in association with a pion in proton-antiproton annihilation
can be seen as the one of the most perspective ways to access $\pi N$ TDAs at \={P}ANDA.
Indeed, historically
\cite{PQCD}
the description of charmonium exclusive decays into hadrons has been one
of the first examples of successful applications of perturbative QCD methods for hard
exclusive reactions.   The annihilation of the
$c \bar{c}$
pair into the minimal possible number of gluons producing quark-antiquark pairs, which then
form outgoing hadrons, was recognized to be the dominant mechanism.
The factorized description of
\begin{equation}
  \bar N (p_{\bar N}) \;+   N (p_N) \; \to  J/\psi(p_{\psi})\;+\; \pi(p_{\pi})
  \label{reac}
\end{equation}
in terms of $\pi N$ TDAs and nucleon DAs
proposed in
\cite{Pire:2013jva}
is based on the extension of the same perturbative QCD framework.
The
$\bar{N} N$
center-of-mass energy squared
$s=(p_N+p_{\bar{N}})^2=W^2$
and the charmonium mass squared
$M_\psi^2$
introduce the natural hard scale for the problem.
It is argued that the reaction
(\ref{reac})
admits a factorized description in the near forward
($t \equiv (p_\pi - p_{\bar{N}})^2 \sim 0$)
and  near backward
($u \equiv (p_\pi - p_N)^2 \sim 0$)
kinematical regimes. The corresponding mechanisms are presented on
Fig.~\ref{fig-0_Semenov}.
Similarly to the case of
$ \bar{N} N$
annihilation into a high invariant mass lepton pair in association with a pion,  
$C$
invariance results in the perfect symmetry between the forward and backward regimes of the reaction
(\ref{reac}).

Below for definiteness we consider the near forward regime. The
$z$
axis is chosen along the colliding
$ \bar{N} N$
with the positive direction along the antinucleon beam. We introduce the light-cone vectors satisfying
$2 p^t \cdot n^t=1$
and define the
$t$-channel skewness variable
$ \xi \equiv - \frac{(p_\pi-p_{\bar{N}}) \cdot n^t}{(p_\pi+p_{\bar{N}}) \cdot n^t} \simeq \frac{M^2_\psi}{2 W^2 -M^2_\psi}$.
Following
\cite{Chernyak:1987nv},
in our calculation we set the relevant masses to the average value
$ M_\psi  \simeq  2m_c   \simeq \bar{M} =3 \; {\rm GeV }$.
The physical kinematical domain for the reaction
(\ref{reac})
in the backward regime is determined by the requirement that the transverse momentum transfer should be space-like:
$(p_\pi-p_{\bar{N}})_T^2 \equiv {\Delta_T^2} \le 0$,
where
$
{\Delta_T^2}= \frac{1-\xi}{1+\xi} \left(t- 2\xi \left[\frac{m_N^2}{1+\xi} -\frac{m_\pi^2}{1-\xi} \right] \right).
$

The leading order amplitude of the reaction
(\ref{reac})
from the mechanism presented on
Fig.~\ref{fig-0_Semenov}
was computed in
\cite{Pire:2013jva}. It reads:
\begin{eqnarray}
{\cal M}^{s_N s_{\bar{N}}}_\lambda
=    {\cal C} \frac{1}{{\bar M}^5 }
\bar{V}(p_{\bar{N}},s_{\bar{N}} ) \Big[ \hat{\cal E}^*(\lambda) \gamma_5  {\cal J}(\xi,t) 
-\frac{1}{m_N}   \hat{\cal E}^*(\lambda) \hat{\Delta}_T \gamma_5   {\cal J}'(\xi,t)
\Big] U(p_N, s_N),
\label{Amplitude_master}
\end{eqnarray}
were 
$U$ 
and 
$V$ 
are the usual Dirac spinors, 
$\hat{\cal E}$ 
denotes the charmonium polarization vector and
${\cal J}, \,{\cal J}'$
stand for the convolutions of the hard kernels with
$\pi \bar{N}$
TDAs and nucleon DAs. The explicit expressions for
${\cal J}, \,{\cal J}'$
worked out in
\cite{Pire:2013jva} 
allow the calculation of these convolutions within specific phenomenological
models for 
$\pi N$ 
TDAs. The normalization factor
${\cal C}$
in
(\ref{Amplitude_master}) reads
${\cal C}=(4 \pi \alpha_s)^3 \frac{f_N^2 f_\psi}{f_\pi}  \, \frac{10}{81}$,
where
$\alpha_s$
is the strong coupling,
$f_\pi=93$
MeV is the usual pion decay constant and
$f_\psi=413 \pm 8$ MeV
is the normalization constant of the heavy quarkonium wave function.
Its value is fixed from the charmonium leptonic
decay width
$\Gamma(J/ \psi \to e^+ e^-)$.

Within the suggested reaction mechanism it is the transverse polarization of charmonium that
is relevant to the leading twist accuracy. Summing over the transverse polarizations of charmonium
and averaging over spins of initial nucleon and antinucleon we get
\begin{eqnarray}
|\overline{\mathcal{M}_{T}}|^2 \equiv \sum_{\lambda_T}  \frac{1}{4}
\sum_{s_N s_{\bar{N}}} {\cal M}^{s_N s_{\bar{N}}}_\lambda \left( {\cal M}^{s_N s_{\bar{N}}}_{\lambda'} \right)^*=
\frac{1}{4} |\mathcal{C}|^2 \frac{2(1+\xi)}{\xi {\bar{M}}^8} 
  \left( |\mathcal{J}(\xi, t)|^2 - \frac{\Delta_T^2}{m_N^2} |\mathcal{J}'(\xi, t)|^2 \right).
\end{eqnarray}
Therefore, the leading twist differential cross section of
$N + \bar{N} \to J/\psi + \pi$
reads
\begin{eqnarray}
\frac{d \sigma}{d t}= \frac{1}{16 \pi \Lambda^2(s,m_N^2,m_N^2) } |\overline{\mathcal{M}_{T}}|^2, \ \ {\rm where} \ \
\Lambda(x,y,z)= \sqrt{x^2+y^2+z^2-2xy-2xz-2yz}.
\label{CS_def_delta2}
\end{eqnarray}

\section{Counting rates studies for \={P}ANDA}
\label{sec-2_semenov}
The dedicated feasibility studies of the reactions admitting description in terms of baryon-to-meson
TDAs  are now coming into the focus of intensive studies
of several experimental groups involved into the  \={P}ANDA project.
The new event generator within the PANDARoot framework was developed by one of us
(Binsong Ma) in close collaboration with Beatrice Ramstein. It allows the estimation of the expected counting
rates for the reaction (\ref{reac}) in the near forward and near backward kinematical regimes.
Below we show the first results for the  near forward regime for
$s = 12.25$ GeV$^2$.
As the model input for the cross section the estimates of
Ref.~\cite{Pire:2013jva}
within the nucleon pole model
\cite{Pire:2011xv}
for
$\pi N$
TDAs are employed.

\begin{figure}
\centering
\includegraphics[width=7cm,clip]{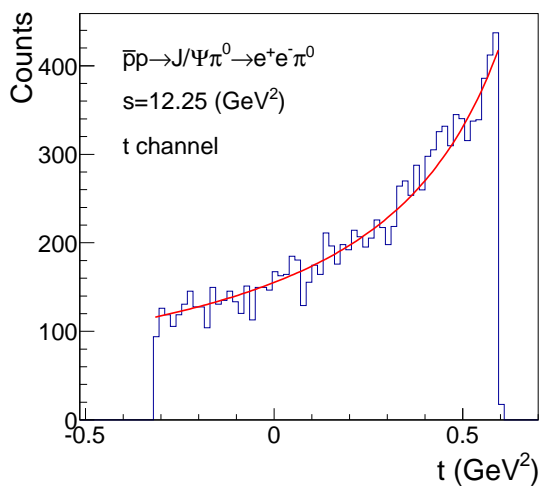}
\caption{Counting rates 
for 
$2$ fb$^{-1}$ 
integrated  luminosity for the near forward regime of the reaction 
(\ref{reac}) 
from the ``PpbarJpsiPi0'' event generator model in the PANDARoot.
The blue histogram shows the results from the generator for bins in 
$t$ 
down from 
$t_{\max}$ 
(corresponding to exactly forward 
$\pi$ 
production). The red line shows the shape of the theoretical prediction from the 
nucleon pole exchange model used as the normalization input for the event generator.}
\label{fig-1_Semenov}       
\end{figure}

With the integrated  luminosity
$2$ fb$^{-1}$
(which corresponds to $4$ months of beamtime at full luminosity)
and assuming for the moment $100 \%$ particle detection efficiency the expected
total number of events is
$N_{\rm tot}=13000$
for near forward
$\pi^0$
emission. Moreover, even better statistics can be expected if one simultaneously
takes into account the information from the backward regime.
On figure~\ref{fig-1_Semenov}
we show the results of studies of counting rates.
The blue histogram shows the results from the generator for bins in
$t$. The bin size is
$0.015$ GeV$^2$.
The red line shows the shape of the theoretical prediction from the nucleon pole exchange model.
The expected large statistics makes us hope for the bright perspectives of accessing 
$\pi N$
TDAs through 
(\ref{reac}) 
at \={P}ANDA.

\subsection*{Acknowledgements}
We would like to thank B. Ramstein for her useful help and advises.
We also acknowledge helpful discussions with T. Hennino, J.P. Lansberg, F. Maas and M. Zambrana.
This work is supported in part by the Polish Grant NCN
No DEC-2011/01/B/ST2/03915,  the Joint Research Activity ``Study of Strongly
Interacting Matter'' (acronym HadronPhysics3, Grant 283286) under the Seventh
Framework Programme of the European Community and by the COPIN-IN2P3 Agreement
and by the Tournesol 2014 Wallonia-Brussels-France Cooperation Programme.

\end{document}